\documentclass[reprint,preprintnumbers,amsmath,amssymb,aip,nofootinbib,prm,onecolumn]{revtex4-2}

\usepackage{xr}
\usepackage[T1]{fontenc}
\usepackage{graphicx}
\usepackage{dcolumn}
\usepackage{bm}
\usepackage{color}
\usepackage{hyperref}
\hypersetup{colorlinks=true,citecolor=red}

\begin{document}

\title{Growth of rhombohedral boron nitride crystals using an iron flux}

\author{W. Desrat$^{1,*}$, M. Moret$^{1}$, J. Plo$^{1}$, T. Michel$^{1}$, A. Ibanez$^{1}$, P. Valvin$^{1}$, V. Jacques$^{1}$, I. Philip-Robert$^{1}$, G. Cassabois$^{1,2}$, B. Gil$^{1}$}
\affiliation{Laboratoire Charles Coulomb (L2C), Universit\'e de Montpellier, CNRS, Montpellier, FR-34095, France}
\affiliation{Institut Universitaire de France, 75231 Paris, France}
\let\thefootnote\relax\footnotetext{* E-mail: wilfried.desrat@umontpellier.fr}

\date{\today}

\begin{abstract}
We report the growth of high quality rhombohedral boron nitride (rBN) crystals by the iron flux method at atmospheric pressure. In contrast to the lamellar structure of standard hexagonal boron nitride (hBN) covering the metal ingot, the current synthetized BN shell is composed of many triangular bulk crystallites, of submillimeter size, predominantly in the rhombohedral phase, with only weak traces of hBN detected by high resolution X-ray diffraction. The low-temperature photoluminescence emission confirms the excellent quality of the rhombohedral stacking with unprecedented width for all phonon replicas at the band edge.
The Raman spectra are characterized by the absence of any low-energy modes at $\sim50$~cm$^{-1}$, the presence of an extended band around $800$~cm$^{-1}$ and an asymmetric high energy mode at $1366$~cm$^{-1}$. These observations are in very good agreement with the phonon dispersion calculated for the rhombohedral primitive cell with $C_{3v}$ symmetry. 
\end{abstract}

\maketitle

\section{Introduction}

Boron nitride (BN) is well known for its remarkable structural, thermal, chemical and optical properties~\cite{Roy2021,Su2024} and its various polymorphs~\cite{Gil2022}. The hexagonal boron nitride (hBN) polytype, which is formed by stacking BN layers with boron and nitrogen atoms arranged in an alternating pattern, has attracted considerable interest due to its large indirect bandgap of approximately $6$~eV. This makes it a promising candidate for use in deep ultra-violet optical devices~\cite{Cassabois2016}. In addition to the well-established AA' stacking (hBN), many other polytypes have been synthesized, including the AB (Bernal) stacking~\cite{Rousseau2021,Rousseau2022} and the AA stacking~\cite{Moon2025}. Each of these structures has different stacking periodicities and crystal symmetries, leading to modifications in their electronic and vibrational properties. Of these atomic arrangements, the rhombohedral phase (rBN), also known as ABC stacking, is of interest due to its non-centrosymmetric nature. This polar material is ferroelectric~\cite{Wang2024} and suitable for nonlinear optics. 
Although it has emerged as a key compound in materials science and nanotechnology, research into its synthesis and characterization is still ongoing~\cite{Zanfrognini2023,Iwanski2024}. The growth of a high-quality macroscopic rBN crystal, with precise measurement of its intrinsic optical and vibrational properties, remains crucial, yet this has not been achieved to date.

The rhombohedral phase of boron nitride crystals was first mentioned by A. H\'erold in 1958~\cite{Herold1958}, who obtained it through the chemical reaction of potassium cyanide with anhydrous borax. The reported lattice parameters of the hexagonal supercell, $a=2.504$~\AA{} and $c=10.01$~\AA{}, indicated a crystalline structure similar to hBN, but with three-layer periodicity along the perpendicular out-of-plane direction. However, the rBN phase was not isolated, but rather mixed with hBN. This indicates that both phases can form simultaneously under identical growth conditions due to similar formation energies~\cite{Gil2022}. Additionally, partial three-dimensional (3D) ordering was observed in the presence of turbostratic stacking~\cite{Thomas1962}, which is also inherent in rBN layers deposited on SiC or AlN/Al$_2$O$_3$ substrates by chemical vapor deposition (CVD)~\cite{Chubarov2011,Olovsson2022}. Turbostratic BN is revealed by a broad X-ray diffraction (XRD) peak on the low-angle side of the main $(003)$ rBN peak, located at $\sim26.3^\circ$~\cite{Chubarov2012,Moret2021}. The synthesis of rBN whiskers with a diameter of the order of $\sim500$~nm by heating commercial hBN powder with oxygen in a graphite crucible~\cite{Ishii1981} suggested that both the oxygen and carbon elements were necessary for rBN formation. This was later refuted, as oxides do not appear to favor rhombohedral stacking, unlike chlorides and cyanides~\cite{Sato1985}. Indeed, chemical syntheses of rBN obtained by heating NaBH$_4$+NH$_4$Cl+KCN above $1000$~$^\circ$C and then annealing the resulting powder at $2100$~$^\circ$C in nitrogen atmosphere provide boron nitride with a high yield of rBN and even pure rBN with no XRD signature of hBN~\cite{Taniguchi1997}. The drawback was the size of the particles, which had a diameter of around $200$~nm. Recent cathodoluminescence and Raman spectra have been reported for these particles~\cite{Zanfrognini2023}. Other techniques have been used to try to grow high-quality rBN, such as ammonothermal methods in an autoclave~\cite{Maruyama2018,Dooley2025}, metalorganic vapor phase epitaxy on AlN buffers~\cite{Iwanski2024} or controlled epitaxy of rBN films on NiFe foils or bevel facets of Ni substrates~\cite{Wang2024,Qi2024}. Despite the very high quality of hBN crystals grown by the metal flux method at atmospheric pressure~\cite{Liu2017} and its ability to produce BN samples with isotopic substitution~\cite{Liu2018}, the synthesis of rBN bulk crystals by this method has not been particularly studied. However, rhombohedral stacking has been identified in a BN sample grown from a nickel and chromium solvent in the presence of carbon~\cite{Rousseau2021}, proving that this technique can lead to rBN stacking. 
  
Here we demonstrate that rhombohedral boron nitride bulk crystallites can be grown from molten iron at atmospheric pressure. Powder X-ray diffraction reveals that these crystallites have excellent crystalline quality, with a rhombohedral phase that dominates the hexagonal phase. This enables us to revisit the physics of this BN polytype using improved optical measurements with Raman and photoluminescence (PL) spectroscopies performed on individual monophasic crystals. The electronic and vibrational properties of rBN are well described using the primitive rhombohedral cell in density functional theory (DFT) simulations.

\section{Rhombohedral boron nitride crystals grown by an iron flux}
\begin{figure}[h]
\includegraphics[width=0.9\columnwidth]{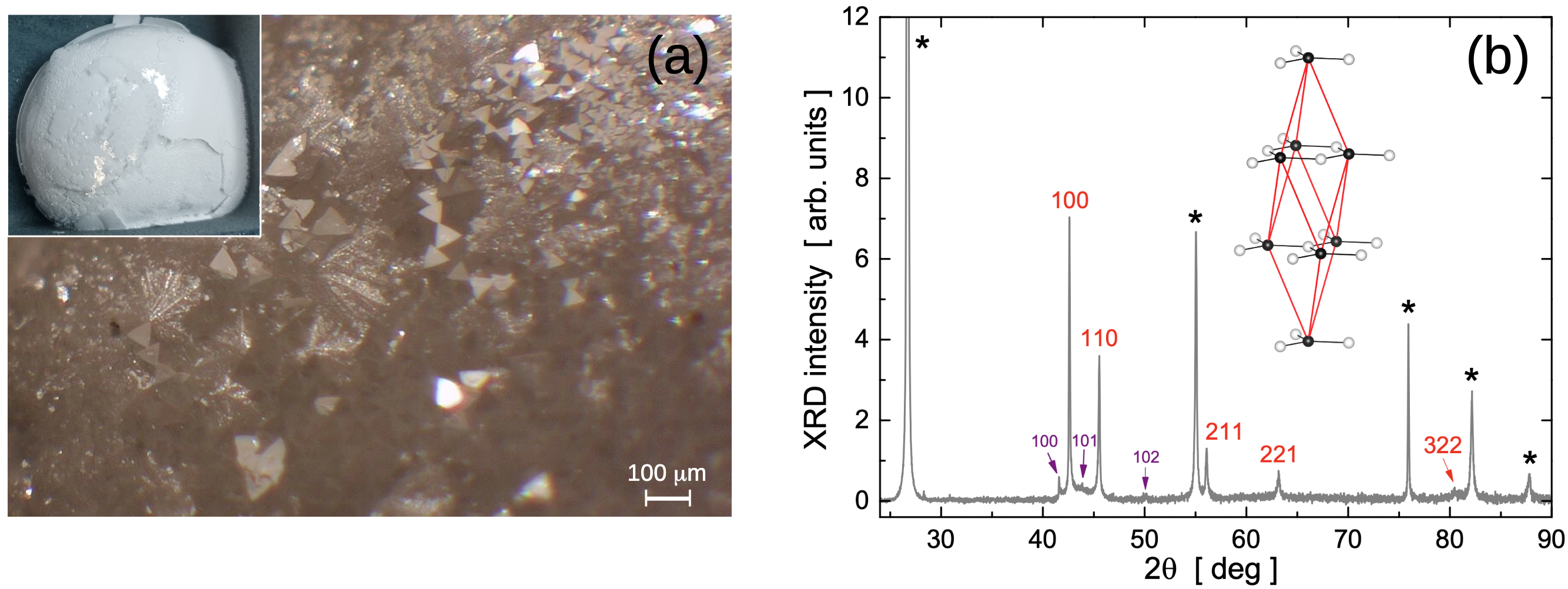}
\caption{\label{fig1} (a) Optical image of the boron nitride shell surface with triangular shaped crystals. The inset is a picture of the as-grown metallic ball covered by white BN. (b) X-ray diffractogram of the BN shell reduced into powder. The red figures identify the rhombohedral lattice planes, the purple figures the hexagonal lattice ones and the star symbols the common diffracting planes. The rhombohedral primitive cell is plotted in inset.}
\end{figure}

Boron nitride samples were grown using the iron flux method at atmospheric pressure, in accordance with the processes detailed in previous studies~\cite{Li2021,Li2021b,Ouaj2023}. Boron powder ($>98\,\%$ purity) and iron powder ($99\,\%$ purity) were mixed at a ratio of $2.5$~wt$\%$ and placed inside an alumina crucible. The mixture was melted in a tubular furnace that had been purged with nitrogen, at temperatures of either $1550^\circ$C or $1600^\circ$C, under a continuous flow of $200$~sccm Ar/H$_2$. The mixture was then left in an N$_2$ atmosphere for $24$~hours to saturate the flux with nitrogen. The pressure was fixed at $1130$~mbar using a needle valve on the exhaust. Boron nitride formed during slow cooling to $1550^\circ$C or $1500^\circ$C at a rate of $-0.02^\circ$C/min or $-0.04^\circ$C/min, followed by a faster decrease to room temperature. This resulted in a white, fluffy shell that covered the entire metal ball (see inset of Fig.~\ref{fig1}(a)), which looked very different from the large, transparent hBN layers that covered the surface of Fe ingots in most previous works~\cite{Li2021,Li2021b,Ouaj2023}. Zooming in reveals triangular-shaped crystallites with a lateral size of several tens of microns at the surface of the shell (Fig.~\ref{fig1}(a)). These crystallites have a flat top surface and are mostly oriented parallel to the metal surface, although crystal tilting is visible in the scanning electron microscopy image (Fig.~S\ref{fig1SM}). The X-ray diffractogram of a millimeter-sized piece of the as-grown shell, measured in a $\theta/2\theta$ configuration with respect to the shell surface (Fig.~S\ref{fig2SM}(a)), exhibits $(00l)$ peaks (or $(111)$ peaks, in rhombohedral notation), identifying the stacked BN atomic planes. Thus the triangular crystals grow preferentially along the $[001]$ axis, or the $[111]$ axis in the rhombohedral coordinate system. This suggests that crystal growth occurs with the boron nitride atomic planes stacked preferentially parallel to the surface of the molten flux. The diffractogram of a powder sample measured with a monochromator is plotted in Fig.~\ref{fig1}(b) with the long-range background removed. Many peaks can be seen, which can be divided into several groups. The peaks of the hexagonal boron nitride phase are labelled in purple, those of the rhombohedral boron nitride phase in red, and star symbols indicate the diffraction angles common to both phases. The rBN peaks are more intense than the hBN ones, reflecting the dominant rhombohedral phase in the samples. More precisely in the $41^\circ-46^\circ$ range, the $(100)$ and $(101)$ hBN peaks are approximately ten times smaller in magnitude than the $(100)$ and $(110)$ rBN peaks. The powder diffraction files $045\mbox{-}1171$ and $034\mbox{-}0421$ for each phase~\cite{icdd} indicate comparable relative values with respect to the main peak, $(002)$ or $(003)$, at approximately $26.7^\circ$. The quantitative analysis detailed in the Supplementary Material indicates that the proportion of the rhombohedral phase is close to $93\,\%$ by weight of the sample. This value is obtained from a macroscopic powder sample resulting from the mixture of hexagonal and rhombohedral crystals, as will be demonstrated later by the local characterization of individual crystallites. Therefore, we can assume that there is a large proportion of rBN crystals in the outer shell, with probably a reduced amount of polytypism, namely hBN, within a single crystallite. The lattice parameters of the rBN phase, as obtained from Rietveld refinement, are $a=2.5041$~\AA{} and $c=9.9981$~\AA.

The diffractogram in Fig.~\ref{fig1}(b) is very similar to that of the rBN powder reported by Sato {\it et al.}~\cite{Sato1985}, which also contained traces of hBN. However, that powder consisted of submicrometer crystallites with an average size of $200$~nm~\cite{Zanfrognini2023}, whereas the triangular crystals grown in this study are up to $100$ $\mu$m in size within the plane and $1$~mm in length along the growth axis. More generally, the best XRD signatures of rhombohedral crystals were obtained from submicrometer triangular platelets~\cite{Xu2007,Bao2009}, whereas rBN films grown by CVD are often degraded by turbostratic stacking, as revealed by the asymmetric $(003)$ peak at around $26.7^\circ$~\cite{Chubarov2012,Moret2021,Olovsson2022}. Figure~S\ref{fig2SM}(b) shows the X-ray diffractograms of three samples milled into powder. Only the slow cooling step differs slightly in terms of setpoint temperatures and cooling rates. All of them demonstrate the dominance of the rhombohedral phase over the hexagonal phase and the reproducibility of the rBN growth. This shows that the iron solvent technique can be used to synthesize high-quality, bulk rhombohedral boron nitride crystallites with no evidence of turbostratic stacking. 

\section{Raman spectroscopy}
\begin{figure}[h]
\includegraphics[width=0.9\columnwidth]{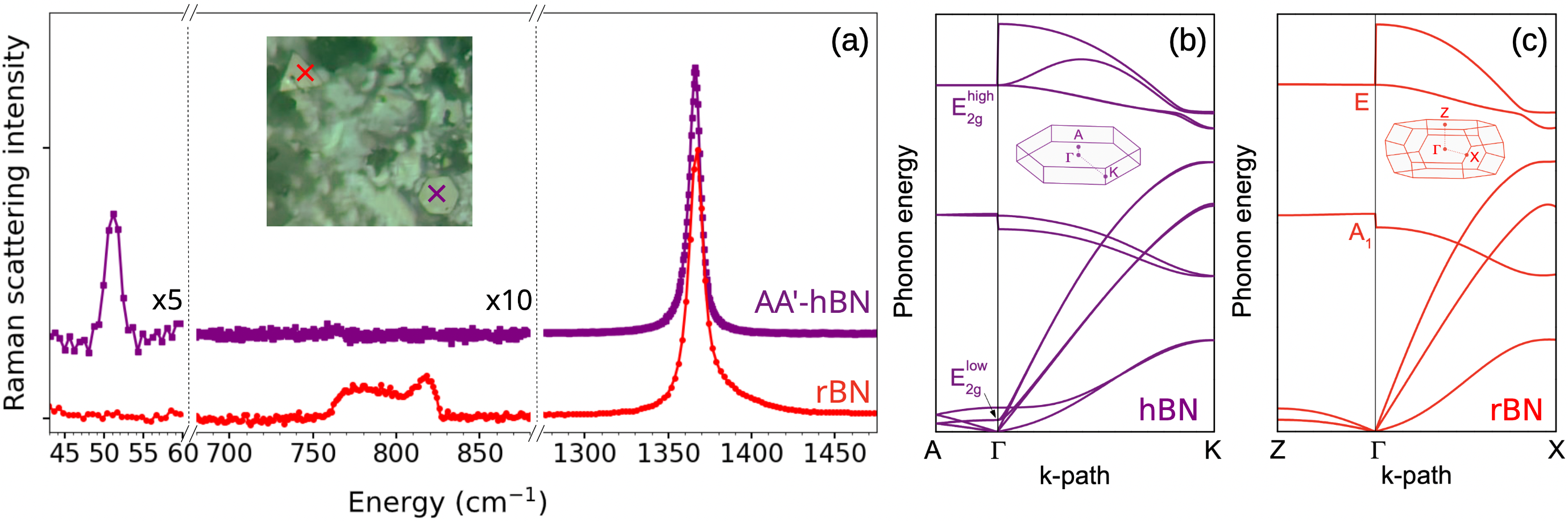}
\caption{\label{fig2} (a) Raman spectra of hBN and rBN measured on hexagonal and triangular crystallites respectively, pictured in the inset. (b) and (c) Phonon dispersions of hBN and rBN around the $\Gamma$ point with the labelled Raman active modes. The Brillouin zones are sketched in inset.}
\end{figure}

In addition to the X-ray diffraction studies performed on the rBN powder, microRaman spectroscopy was used to identify individual crystallites based on their lattice vibration modes. Figure~\ref{fig2}(a) shows the Raman spectra measured in a backscattering configuration on the $(001)$ planes of two independent crystallites: one triangular and one hexagonal, as can be seen in the inserted optical image. The upper spectrum, obtained from the hexagonal crystal, is characteristic of the hBN phase~\cite{Janzen2024}, with AA' stacking. This consists of two symmetrical peaks : one at low frequency ($51$~cm$^{-1}$), of low intensity, and another which dominates at high frequency, centered at $1365$~cm$^{-1}$ with a width of $8.4$~cm$^{-1}$. These are the two active Raman modes $E_{2g}^\text{low}$ and $E_{2g}^\text{high}$ at the $\Gamma$ point, corresponding to in-plane vibrational modes. Indeed the hBN crystal of $D_{6h}$ symmetry possesses 12 vibration modes, of which 4 are Raman-active optical modes, the $E_{2g}^\text{low}$ and $E_{2g}^\text{high}$, which are double degenerate at $k=0$. These modes are indicated in the phonon dispersion of hBN plotted vs the $A-\Gamma-K$ path in Fig.~\ref{fig2}(b). The phonon dispersions were computed for illustrative purposes and do not include the excitonic or van der Waals corrections that are considered in the dedicated Density Functional Theory (DFT) calculations~\cite{Sponza2018,Zanfrognini2023,Mishra2025}.

The lower spectrum in Fig.\ref{fig2}(a) differs from the previous one in three main ways: (i) the low-frequency mode is not detected despite the same measurement conditions, (ii) a broad band appears between $760$~cm$^{-1}$ and $825$~cm$^{-1}$, and (iii) the high-frequency mode is still present, but with slight asymmetry resulting in a tail on the high-energy side. The suppression of the low-energy mode and the appearance of a mode around $800$~cm$^{-1}$ are features typical of rhombohedral BN and have already been reported~\cite{Sato1985}. However, the shape and width of the intermediate band and the asymmetry of the band at $1365$~ cm$^{-1}$ have never been explicitly noticed before, although they have been observed in several rBN spectra~\cite{Liu1995,Zanfrognini2023,Dooley2025} and therefore seem to be inherent characteristics of this material. Before discussing the possible origin of these features, it is worth emphasizing that the detected modes are in perfect agreement with the phonon modes calculated for the rhombohedral lattice and plotted along $Z-\Gamma-X$ in Fig.~\ref{fig2}(c). Unlike many phonon mode calculations based on a non-primitive hexagonal cell containing six atoms - a direct representation of the threefold ABC stacking of BN layers - the phonon dispersion shown in Fig.~\ref{fig2}(c) has only six modes. In addition to the three acoustic modes, rBN with $C_{3v}$ symmetry exhibits three optical modes, which are both Raman and infrared active. The out-of-plane $A_1$ mode occurs at around $800$~cm$^{-1}$ and the in-plane double-degenerate $E$ modes occur at $1365$~cm$^{-1}$ at the center of Brillouin zone, as in hBN. However, the absence of the lower LO$_1$, TO$_1$ and ZO$_1$ branches in the $k_z=0$ plane is notable. Consequently, the $E_{2g}^\text{low}$ shear mode does not exist at the $\Gamma$ point in rBN, which explains why the $E_{2g}^\text{low}$ mode is not observed in the Raman spectrum. This is due to the presence of only two atoms in the primitive rhombohedral cell of the rBN crystal, the corresponding Brillouin zone of which is sketched in the inset of Fig.~\ref{fig2}(c). The dispersion of rBN with six modes had previously been calculated by Yu {\it et al.}~\cite{Yu2003}. However, widespread phonon dispersions computed for the ABC stacking consist of 18 phonon modes, 12 of which result from band folding for non-zero $k_z$. A more detailed discussion is provided in the Supplementary Material. 

On average, the $E$ mode is centered at $1366$~cm$^{-1}$, with a full width at half maximum (FWHM) of only $10$~cm$^{-1}$. This is slightly wider than in bulk hBN, but still confirms the very good crystalline quality of the rBN crystallites. The weak asymmetry of the peak is consistently present at its base on the high-energy side. The band around $800$~cm$^{-1}$, as shown in Fig.\ref{fig2}(a), is typical. It has a non-constant amplitude and appears to be made up of several components, including a distinct peak around $820$~cm$^{-1}$. Figure~S\ref{fig8SM} presents a series of spectra obtained from various rBN crystallites, which reveals that different shapes can form. The band spread remains constant, between $760$~cm$^{-1}$ and $825$~cm$^{-1}$ with abrupt flanks on either side. This is unlike the Lorentzian lineshape of the $E$ mode. The various band shapes detected with different positions of the maximum probably explain the dispersion of the reported experimental energies for the $A_1$ Raman mode in rBN, which ranges from $790$~cm$^{-1}$ to $800$~cm$^{-1}$~\cite{Liu1995,Sato1985,Zanfrognini2023}.  

It is interesting to compare the limits of the spread band, $760$~cm$^{-1}$ and $825$~cm$^{-1}$ with the two phonon modes $A_{2u}(TO)=767$~cm$^{-1}$ and $A_{2u}(LO)=825$~cm$^{-1}$, which were evaluated experimentally by Segura {\it et al.} using infrared (IR) reflectance measurements on hBN~\cite{Segura2019}. It should be noted that the $A_1$ and $E$ modes of rBN, which is non-centrosymmetric (space group {\it R3m}) unlike hBN, are both Raman and IR active. The $A_{2u}$ energies are in good agreement with the extended band detected in Raman spectroscopy, which can be associated with the Restrahlen band of the out-of-plane mode ($A_{2u}$ for hBN and $A_1$ for rBN). The values of the $A_1$ mode computed by DFT at the $\Gamma$ point lie between $760-770$~cm$^{-1}$~\cite{Zanfrognini2023,Yu2003}, which suggest that the energy of the $A_1$ mode should coincide with the lower bound of the detected Raman band. Recently, a significant correction shift in the energy of the $A_1$ mode was demonstrated for the ABC stacking, from $778.5$~cm$^{-1}$ to $730.6$~cm$^{-1}$, due to van der Waals interactions between adjacent layers~\cite{Mishra2025}. However, this last value seems an underestimation compared to the experimental range of the Raman signal. Similarly, the high-energy tail of the $E$ peak could be linked to the Restrahlen band of the in-plane mode, which extends from $E_{1u}(TO)=1365$~cm$^{-1}$ up to $E_{1u}(LO)=1623$~cm$^{-1}$ in hBN~\cite{Segura2019}. This phenomenon will be investigated through further research on single rBN crystallites coupled to IR spectroscopy.

\section{Low temperature micro-photoluminescence}

\begin{figure}[h] 
\includegraphics[width=0.5\columnwidth]{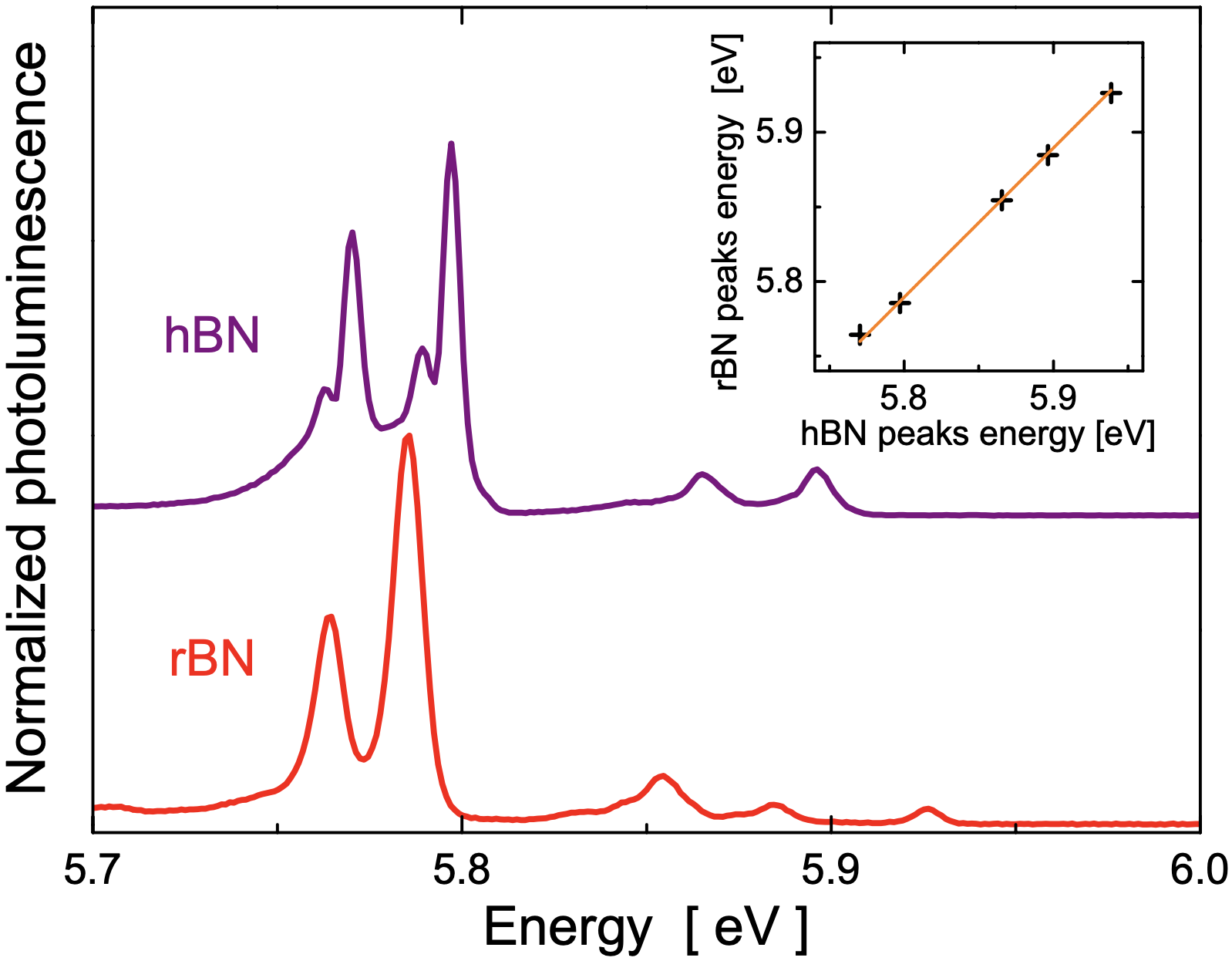}
\caption{\label{fig3} Normalized micro-photoluminescence spectra measured on individual rBN and hBN crystallites at $T=8$~K. The inset plots the major peak positions in rBN vs those in hBN. The offset equals $-10.5$~meV with a unity slope.}
\end{figure}

Figure~\ref{fig3} shows the micro-photoluminescence (PL) spectra of hBN and rBN crystallites at $T=8$~K, focusing first on the phonon-assisted transitions. Within this $5.7-6.0$~eV energy range, the hBN polytype exhibits four primary emission peaks, at $5.896$~eV and $5.865$~eV, for the TA and LA phonon contributions, and at $5.797$~eV and $5.770$~eV, for the TO and LO phonon contributions. The phonons involved in the emission process have a non-zero in-plane wavevector corresponding to the $T$ point of the Brillouin zone (BZ), i.e. half of the $\Gamma K$ distance, ensuring wavevector conservation in accordance with the indirect bandgap. The latter TO/LO peaks present a weaker structure on the low-energy side, with an energy splitting of $\sim7$~meV, corresponding to the $E_{2g}^{low}$ phonon mode at the center of the BZ~\cite{Vuong2017}. These four main lines are also present in the rBN PL spectrum but they are all red-shifted. The inset in Fig.~\ref{fig3} plots the PL peak positions of rBN as a function of the positions measured in hBN. An average shift of $10.5$~meV is estimated, which is slightly smaller than the $15$~meV value reported in Refs.~\cite{Schue2017,Zanfrognini2023}. This red shift may result from either a smaller indirect bandgap or a global shift of the phonon energies in rBN. However, it is evident from the phonon dispersions presented in Ref.~\cite{Zanfrognini2023} that the computed energy difference between the vibrational modes in rBN and hBN is significantly lower than this value, meaning it cannot solely account for the line shift. 
Simply fitting the energy scale of the PL spectrum to that of the phonon dispersion plotted along $\Gamma X$ in the rhombohedral BZ, equivalent of $\Gamma K$ for hBN (see Fig.~S\ref{fig7SM}) results in good agreement between the PL peaks and the energies at half of $\Gamma X$. However, this is obtained for an indirect bandgap fixed at $5.949$~eV, which is only $6$~meV smaller than the indirect bandgap of hBN, which was evaluated to $5.955$~eV~\cite{Cassabois2016}. This discrepancy in shift values suggests that a more in-depth analysis is needed to precisely determine the locations of the conduction band minimum and valence band maximum in the BZ, as well as the wavevectors involved in the indirect optical transition in rBN. DFT calculations predicted a reduction in the bandgap of rBN compared to hBN~\cite{Sponza2018}, but this was overestimated ($\sim500$~meV), since the experimental bandgap difference between the two polytypes is negligible ($\leq10$~meV). Previously, we showed that the rhombohedral primitive cell of rBN allows to reinterpret the Raman data. Similarly, attributing the five main PL peaks to the phonon dispersion shown in Fig.~S\ref{fig7SM}, composed of only six modes, simplifies previous analyses. The PL emission line at $5.926$~eV originates from the low-energy ZA mode, which is permitted in rBN but not in the centrosymmetric hBN. This transition has an extremely weak signature in the hBN PL spectrum (not visible in Fig.~\ref{fig3}) at an energy of $5.939$~eV. 

Thanks to the high cristalline quality of the rhombohedral boron nitride grown by the iron flux, it is worth noting the narrow PL lines. The TO/LO phonon-assisted transitions have a FWHM of $10\pm0.3$~meV, which is much lower than that recorded in previous PL data on powders or CVD layers of rBN. This is only slightly higher than the FWHM of $6.5$~meV in hBN. Thus, the perfectly symmetric LO/TO peaks allows us to conclude that there are no additional emission lines on their low-energy side. This is a direct consequence of the absence of the interlayer shear mode at $\sim50$~cm$^{-1}$ in rBN.

\begin{figure}[h]
\includegraphics[width=0.7\columnwidth]{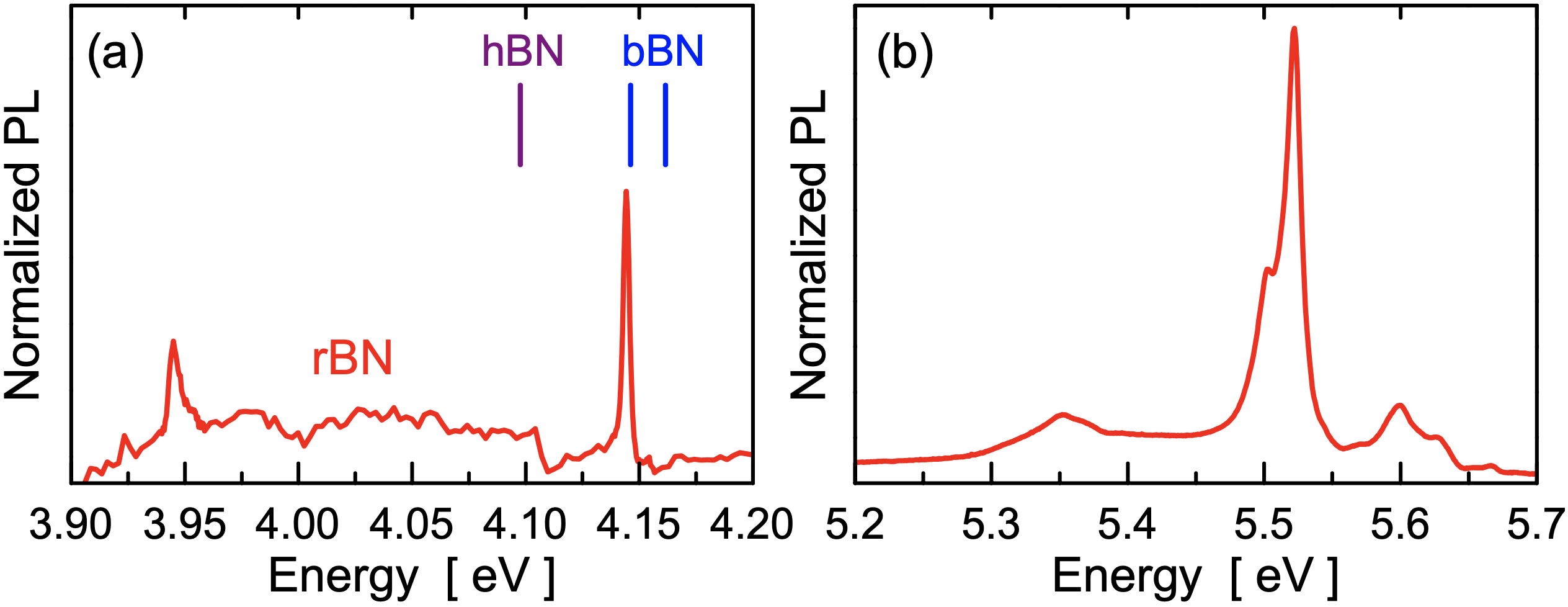}
\caption{\label{fig4} Normalized micro-photoluminescence spectra of rBN in the $3.9-4.2$~eV (a) and $5.2-5.7$~eV (b) ranges, showing the emission peaks from point defects and structural defects respectively. In (a), the vertical lines indicate the energies of the ZPL of the carbon dimer in hBN and bBN polytypes.}
\end{figure}

\begin{figure}[h]
\includegraphics[width=0.5\columnwidth]{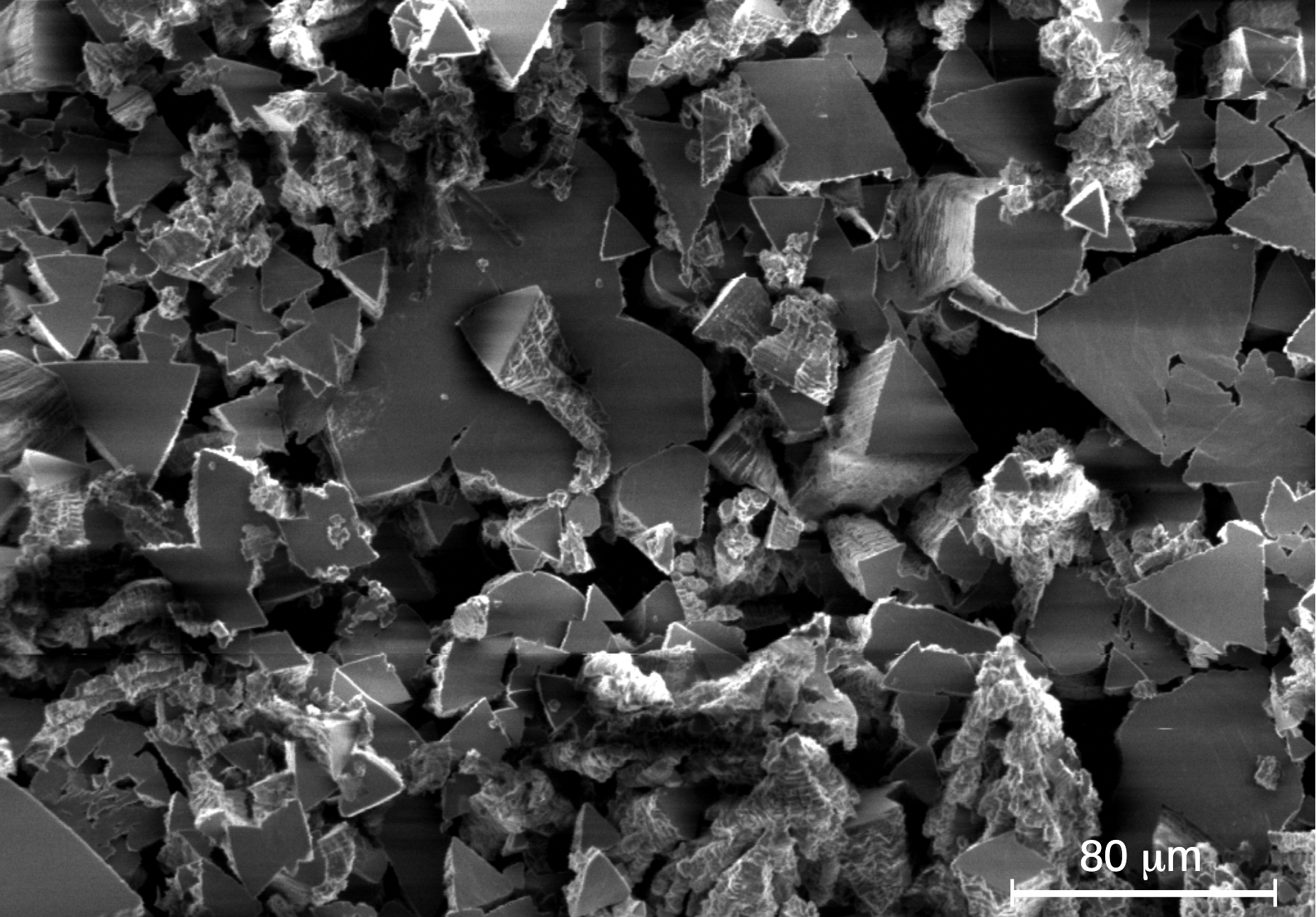}
\caption{\label{fig1SM}  Scanning electron microscope image of the surface of the BN shell covering the iron ingot.}
\end{figure}

\begin{figure}[h]
\includegraphics[width=0.8\columnwidth]{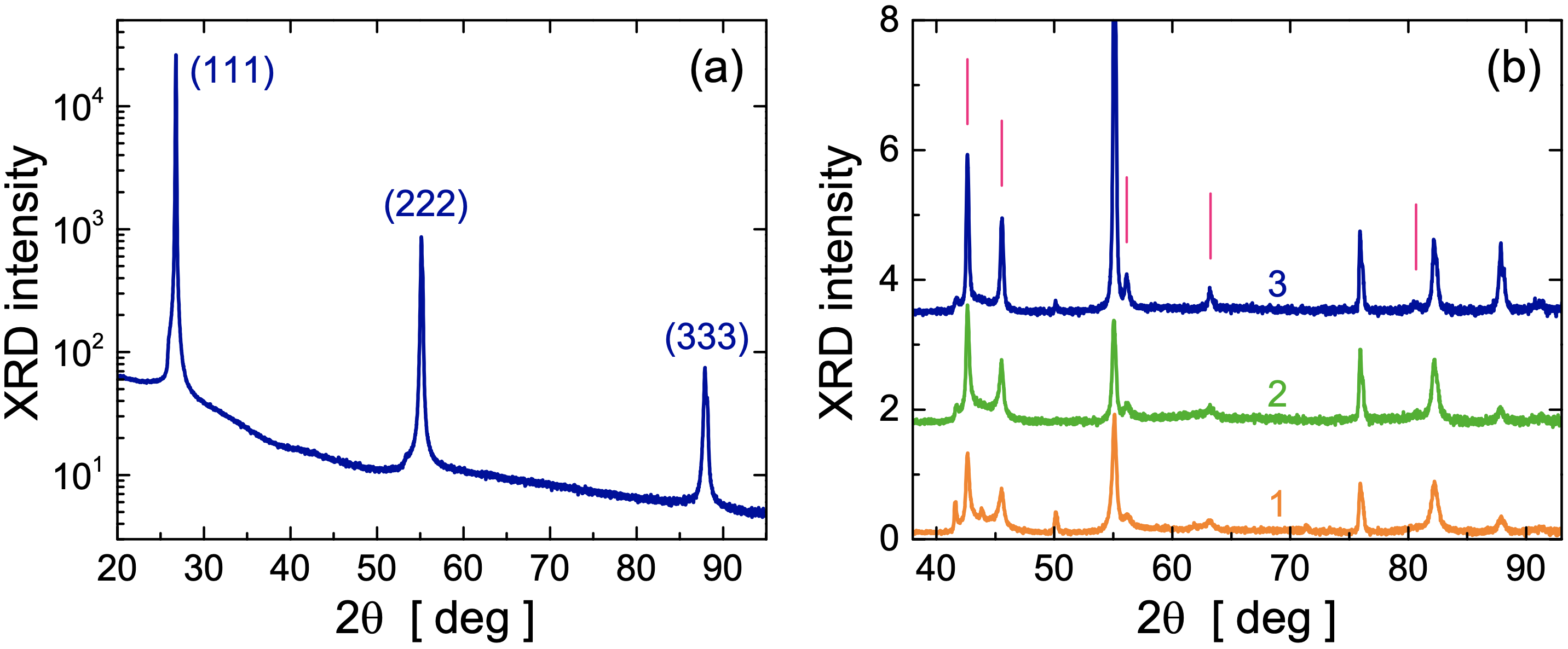}
\caption{\label{fig2SM} (a) X-ray diffractogram of a BN shell with the beam incident to the sample surface in the $\theta/2\theta$ configuration. The peaks identify the stacked BN atomic planes, here labelled with the rhombohedral notation. (b) X-ray diffractograms of 3 powdered BN samples grown with the following parameters for the slow cooling step: (1) $1550^\circ$C to $1500^\circ$C at $-0.02^\circ$C/min, (2) $1600^\circ$C to $1550^\circ$C at $-0.02^\circ$C/min, (3) $1600^\circ$C to $1500^\circ$C at $-0.04^\circ$C/min. Vertical lines point out the rBN specific peaks.}
\end{figure}

Figure~\ref{fig4}(a) focuses on the zero phonon line (ZPL) of the carbon dimer (C2 defect), which is located at $4.1442$~eV for the rBN crystallite, close to the recently reported value of $4.143$~eV~\cite{Iwanski2024}. This is blue-shifted with respect to hBN, where the ZPL stands at $4.0976$~eV. In Bernal BN (bBN or AB stacking), a double structure emerges at $4.1461$~eV and $4.1615$~eV, arising from the double atomic configuration of the C2 defect~\cite{Plo2025}. The near-equality of the ZPL energy in rBN and the lowest ZPL in bBN stems from the carbon dimer's similar environment in both polytypes, with only two nearest-neighbour atoms in adjacent layers~\cite{Iwanski2024}. The blue shift of the ZPL is around $46$~meV between rBN and hBN. Recently, V$_\text{B}^-$ and B-center spin defects were created in rBN flakes and the ZPL of the B-center, which was detected at $2.87$~eV, was found to be blue-shifted by $42.5$~meV compared to hBN~\cite{Gale2025}. This is another proof that point defects are effectively sensitive to stacking orders and that crystalline homogeneity throughout the sample thickness is crucial for narrow emission lines. 
The ZPL intensity measured in the rBN crystallite is weak as can be seen from the noisy background in Fig.~\ref{fig4}(a). This suggests that carbon was incorporated at low levels during the growth of the rBN crystals by the iron flux. Despite the reduced signal intensity, the dominant replica was detected at a lower energy of $3.9450$~eV, leading to an energy splitting of $\Delta=199.2$~meV with the ZPL. The splitting is identical to that measured in hBN within the experimental precision, confirming that the local vibration mode of the carbon dimer in the BN layer has a comparable energy in both polytypes~\cite{Plo2025}. 

Finally, we focus on the intermediate energy range of the PL, in which transitions induced by structural defects such as dislocations or stacking faults occur (see Fig.~\ref{fig4}(b)). There is a well-defined emission peak at $5.522$~eV, which is surrounded by weaker lines at $5.353$~eV, $5.599$~eV, $5.626$~eV and $5.666$~eV. Previous photoluminescence and cathodoluminescence spectra were dominated by broad lines, which prevented the precise determination of their energies~\cite{Xu2007,Zanfrognini2023,Iwanski2024}. For example, PL measurements on rBN layers deposited by CVD were dominated by two bands centered at $5.35$~eV and $5.55$~eV, which were broadened by crystal inhomogeneities or low-energy carrier recombination traps~\cite{Gil2022}. The so-called D-lines are not located at the same energy in rBN as in hBN, and their interpretation in terms of intervalley scattering~\cite{Cassabois2016b} will require further development when the rhombohedral Brillouin zone is considered. 

\section{Discussion}

The results presented above reveal two key features of the crystals obtained using the iron flux technique. Firstly, the samples are small crystals stretched along the $c$-axis. This method is usually known to produce large areas of lamellar and transparent hBN crystals~\cite{Hoffman2014,Liu2018,Ouaj2023}, which are far from the white shell visible in the insert of Fig.~\ref{fig1}(a). However, the white appearance is stipulated in many rBN syntheses~\cite{Sato1985,Xu2007,Bao2009,Dooley2025}. For the solvent method, it has been demonstrated that temperature influences the crystal growth mechanisms, with lower temperatures favoring vertical growth (along the $c$-axis). In the current study, the temperature was fixed at $1550-1600^\circ$C, which is equivalent to the standard temperatures used for the formation of hBN. The slow cooling rate was also set to a typical value of $-0.02^\circ$C/min. Therefore, these growth parameters do not appear to be critical, and neither the pressure nor the nitrogen flow rate seem to be responsible for the aspect ratio of the BN crystals. Initially, the pressure was controlled using a needle valve, resulting in a gradual increase in pressure over time as material accumulated on the needle. To ascertain whether this increase in pressure could be responsible for the growth of crystallites, the pressure was subsequently regulated using a pressure valve controlled by an MKS 600 pressure controller at $1250$~mbar throughout the growth process, rather than $1150$~mbar. However, the formation of standard lamellar crystals suggests that pressure in this range is also not a determining factor. 

Secondly, we emphasized that the rhombohedral polytype is more prevalent than the hexagonal polytype. The rBN stacking had only been detected in BN crystals synthezised by the flux method using a Ni+Cr mixture in the presence of carbon~\cite{Rousseau2021}. However, the rBN polytype was revealed only in powder XRD and not in micro-photoluminescence, which identified Bernal stacking with the typical PL signature at $6.032$~eV. We also observed that, by carrying out a spatial mapping on certain lamellar samples identified as hBN, zones with an rBN signature in PL or Raman are likely to be due to localized stacking defects. In such samples, the detection of the rBN signature is random and difficult to identify because of the recurrent polytypism in these lamellar crystals. Here, the large number of triangular crystallites and the predominance of the rhombohedral polytype make the detection of rBN-specific PL or Raman spectra easier and of better quality. Powder X-ray diffraction also indicates hBN stacking, which was probed by micro-PL on a hexagonal-shaped crystallite. While we cannot rule out polytypism within a single crystallite, the PL spectra obtained by micro-photoluminescence suggest an rBN stacking homogeneity extending several tens of nm in thickness. 

Previously, the presence of certain elements, such as oxygen and carbon, has been cited as favoring the growth of rBN~\cite{Ishii1981}. The iron and boron powders used in this work, with $99\,\%$ and $>98\,\%$ purities, contain traces of numerous elements (Mg, K, F, Si, Ca, S, Ti, etc) that were easily detected by energy dispersive spectroscopy. Titanium was also found in oxide form on the surface of the final BN shell. Therefore, we can infer that impurities in the molten solvent may have favored the nucleation of rhombohedral BN crystals initially. However, after growth, the presence of extrinsic elements in the bulk BN is not detected, which indicates that the impurities are not incorporated during BN crystallization. In this regard, it has recently been demonstrated that rBN grows preferentially on the NiCr walls of an autoclave rather than on cBN seeds~\cite{Dooley2025}. 

Finally, it should be noted that the triangular shape of the crystals seems to be an inherent property of rhombohedral boron nitride, as many rBN platelets have been reported to be triangular, and not hexagonal~\cite{Xu2007,Bao2009,Maruyama2018}. Triangular islands have also been detected in turbostratic BN layers deposited by CVD on AlN/Al$_2$O$_3$ or ZrB$_2$ substrates~\cite{Chubarov2015a,Souqui2021}. First, a layer-by-layer growth mode was observed up to a critical thickness, which is likely linked to stress relaxation arising from the lattice mismatch. This is then followed by an island growth mode~\cite{Chubarov2015a,Chubarov2015b}. In the case of the solvent method, the first layers of boron nitride crystallize on top of molten iron. Then, the growth of the crystals should take place from the bottom, where the solvent is supersaturated with boron and nitrogen, as the temperature decreases. In this scenario, the top layer of the crystallites which appear completely flat in scanning electron microscopy images, would be the first crystalized zones, with growth occurring preferentially along the $c$-axis. As the process continues and the nitrogen content in the solvent diminishes, lateral growth could become more dominant, covering the entire surface of the flux. 

\section{Conclusion}

The synthesis of boron nitride using the iron solvent method at atmospheric pressure is known to produce high-quality hexagonal BN crystals. This work demonstrates that this process is more versatile, enabling the growth of sub-millimeter triangular-shaped crystallites in the rhombohedral phase. Compared to previous epitaxial films, nano-platelets and powders of rBN characterized by concomitant turbostratic stacking phase or broad photoluminescence emission lines, our results show that this technique can significantly improve the quality, homogeneity and size of the crystals. Further investigation is needed to explain the origin of the extended band from $760$~cm$^{-1}$ to $825$~cm$^{-1}$, instead of the expected single $A_1$ mode, in Raman spectra, and to define the indirect bandgap and phonon-assisted transitions in agreement with the rhombohedral Brillouin zone. The small signal of the carbon dimer emission indicates weak carbon incorporation into the BN lattice. This, when associated with controlled rhombohedral stacking, will be ideal for creating and studying point defects, such as spin-based quantum sensors.

\section{Methods}
\subsection{X-ray diffraction}
The X-ray diffractograms are recorded at room temperature in the $\theta/2\theta$ configuration using a Bruker D8 Discover diffractometer fitted with a V\AA{}NTEC linear detector. To enhance detectivity, a G\"obel mirror was used, resulting in a non-monochromatic primary beam composed of the Cu(K$_{\alpha1}$+K$_{\alpha2}$) lines. A nickel filter is inserted to remove the Cu(K$_{\beta}$) line. When mentioned, a monochromator is inserted to ensure only the Cu(K$_{\alpha1}$) emission is present.

\subsection{Raman spectroscopy}
Micro-Raman spectra were measured at room temperature using a Renishaw Invia spectrometer (2400 lines per mm) from $500$~cm$^{-1}$ to $1600$~cm$^{-1}$ with either a $532$ or $785$~nm laser. The low-energy mode at $\sim50$~cm$^{-1}$ was measured using a Horiba T64000 (1800 lines per mm) with a $660$~nm laser in subtractive mode. For this, we employed the oversampling method described in Ref.~\cite{Vuong2018}. In both setups, the laser was focused using a $50\times$ objective (N.A. $0.5$) with the power kept below $300$~$\mu$W to prevent the sample from heating up.

\subsection{Photoluminescence}
Photoluminescence spectroscopy is carried out using an excitation beam provided by the fourth harmonic of a continuous-wave mode-locked Ti:Sa oscillator. This oscillator is tunable from $193$~nm to $205$~nm, delivering trains of $140$~fs pulses at a repetition rate of $80$~MHz. The spot diameter is approximately $240$~nm with a power of $110$~$\mu$W. The exciting laser beam, tuned at $197.4$~nm, is focused by a Schwarzschild objective located inside a closed-cycle cryostat equipped with CaF$_2$ optical windows. The detection system comprises a Czerny-Turner monochromator with a focal length of $500$~mm, fitted with a $1200$~grooves/mm grating blazed at $250$~nm and a back-illuminated charge-coupled device camera (Andor Newton 920), with a quantum efficiency of $50$~$\%$ at $210$~nm. The sample is mounted on a piezoelectric stepper and scanner assembly, which is cooled down to $6$ K under an ultrahigh vacuum of $10^{-8}$~mbar.

\section{Acknowledgments}
 The authors acknowledge the support of the Occitanie Region through the D\'efi Cl\'e "Technologie Quantique" initiative.

\section{Conflict of Interest}
The authors declare no conflict of interest.

\section{Data Availability Statement}
The data that support the findings of this study are available from the corresponding author upon reasonable request.

\bibliography{arxiv_main}

\newpage
\section*{SUPPLEMENTARY MATERIAL}

\subsection{Rhombohedral boron nitride crystals and X-ray diffraction}

\begin{figure}[h]
\includegraphics[width=0.6\columnwidth]{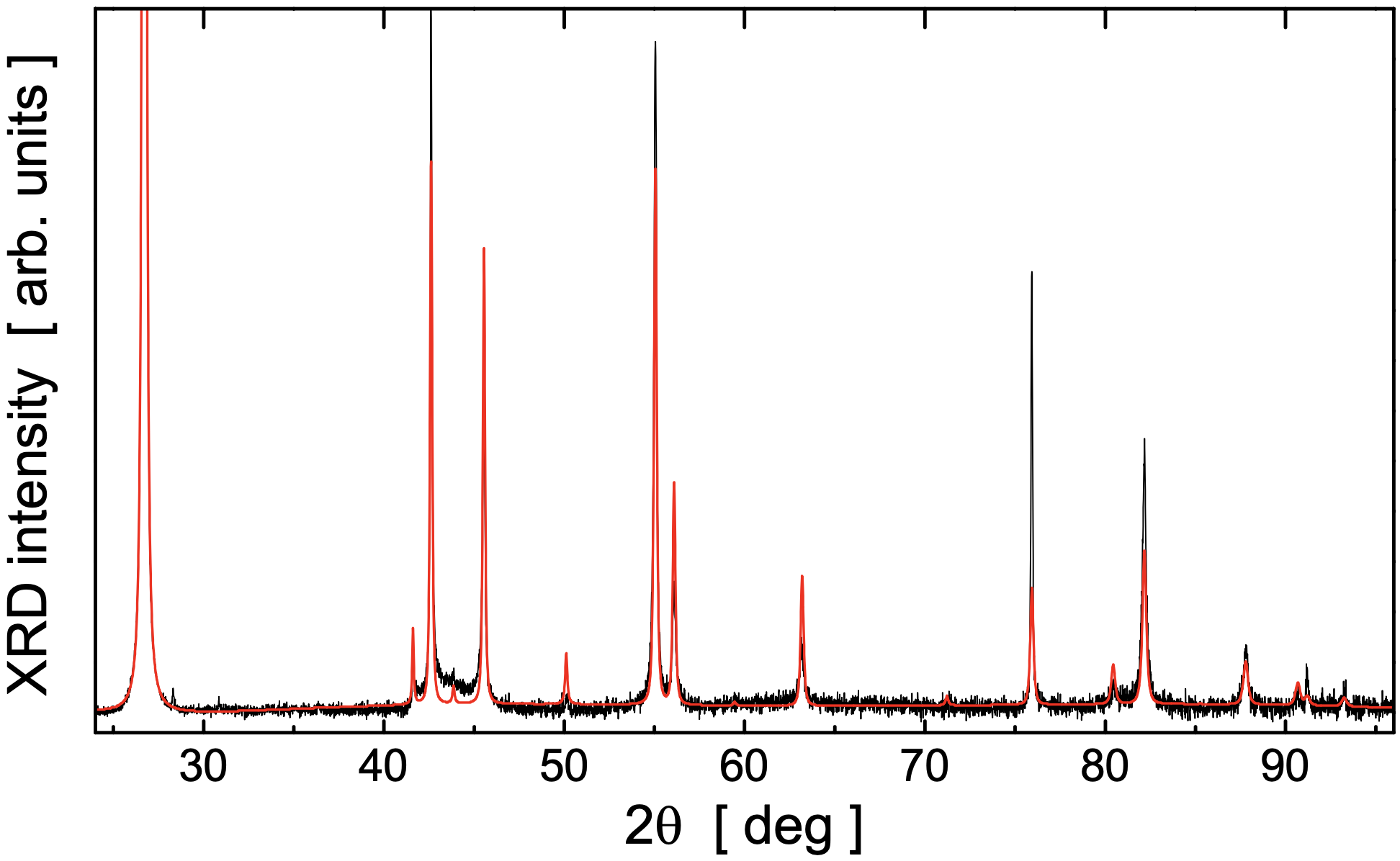}
\caption{\label{fig2SMc} Experimental powder X-ray diffractogram (black curve) modeled by Rietveld refinement (red curve) with the Profex software (https://www.profex-xrd.org), which included isotropic micro-strain, anisotropic crystallite size and preferential crystal orientation. A weight ratio of $92\,\%$ rBN is obtained.}
\end{figure}

\subsection{Hexagonal and rhombohedral cells of ABC-stacked boron nitride}
\begin{figure}[h]
\includegraphics[width=0.4\columnwidth]{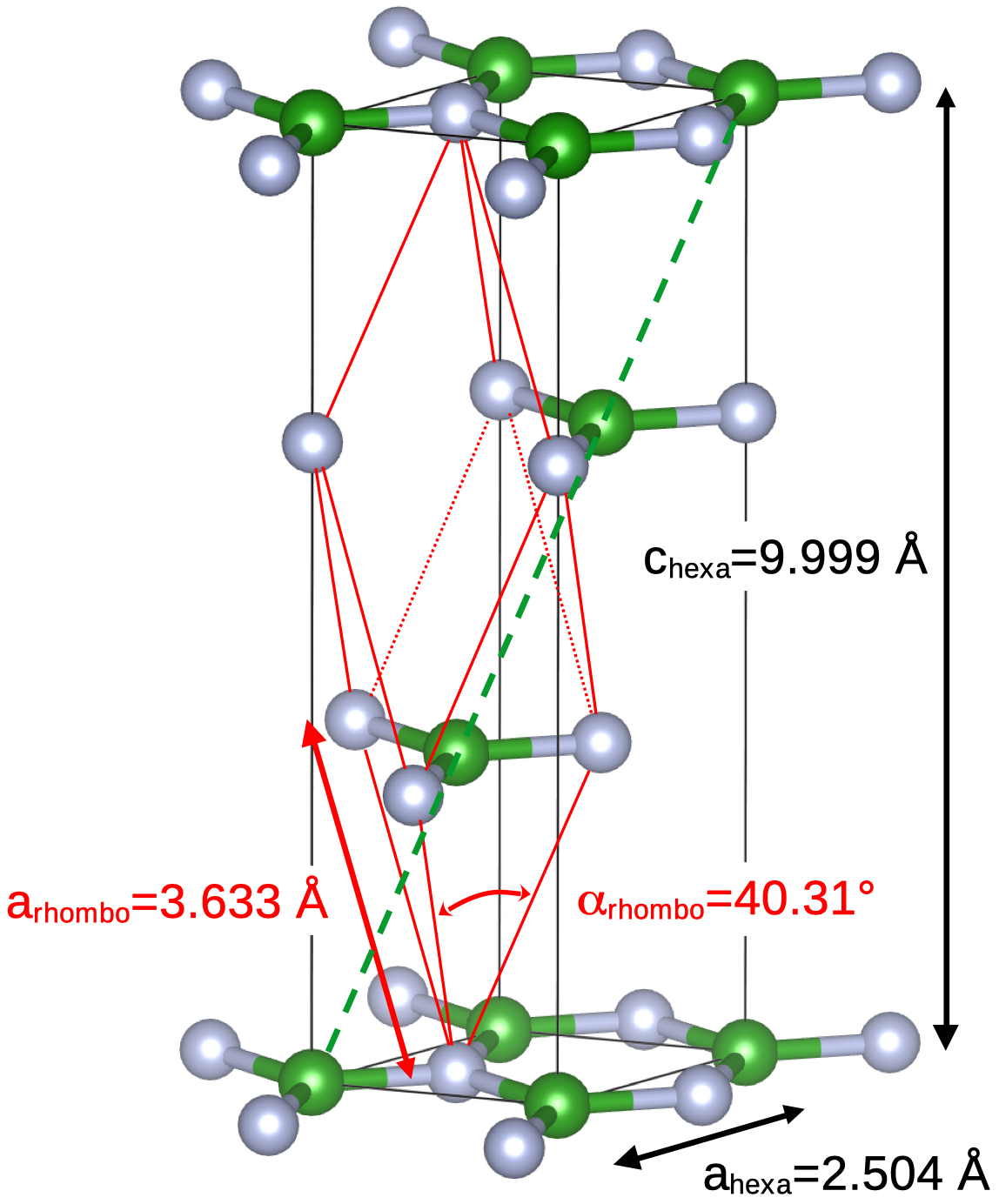}
\caption{\label{fig3SM} Hexagonal cell (black line) and rhombohedral cell (red line) of the ABC-stacked boron nitride. The green dashed line is a guide to the eye of the translation of adjacent atomic planes.}
\end{figure}

The ABC polytype of boron nitride consists of a periodic stacking of three BN atomic planes along the $[001]$ direction (Fig.~\ref{fig3SM}). Each BN layer shifts in-plane by the bond length along a BN bond with respect to its neighbors (illustrated by the oblique green dashed line in Fig.~\ref{fig3SM}). ABC boron nitride is classically treated as a hexagonal crystal cell because of the simple identification of the Miller indices or the definition of the $k$-path in the reciprocal space, thanks to the consequent hexagonal Brillouin zone. In reality, however, the ABC stacking corresponds to the rhombohedral Bravais lattice. Considering this primitive cell in the electronic properties calculation reduces the number of bands in the energy diagram and in the phonon dispersion compared to the hexagonal cell approach, allowing the experimental results to be revisited.

Figure~\ref{fig3SM} displays the ABC-stacked boron nitride with the conventional hexagonal cell in black and the primitive rhombohedral cell in red. The hexagonal cell is defined by two in-plane vectors of equal size ($a_\text{hexa}\approx2.504\,\AA$), separated by the angle $\gamma=120^\circ$, and a third vector along the $z$-axis (i.e. $\alpha=\beta=90^\circ$), which is $3$ times the interplanar distance ($c_\text{hexa}\approx9.999\,\AA$). The rhombohedral cell is formed by three vectors of equal size ($a_\text{rhombo}\approx3.633\,\AA$), which are separated by equal angles, $\alpha=\beta=\gamma\approx40.31^\circ$, forming a triangular pyramid with a $C_{3v}$ symmetry along the $z$ axis.
The parameters of the rhombohedral cell can be easily deduced from the hexagonal lattice parameters, with $a_\text{rhombo}=\sqrt{(a_\text{hexa}/\sqrt{3})^2+(c_\text{hexa}/3)^2}$ and
$\alpha_\text{rhombo}=\arccos(-\frac{1}{2}\sin^2\theta+\cos^2\theta)$ where $\theta=\arctan(a_\text{hexa}\sqrt{3}/c_\text{hexa})$.

The hexagonal cell has six atoms in the positions $(0,0,0)$, $(2/3,1/3,1/3)$, and $(1/3,2/3,2/3)$ for one element (B or N) and $(2/3, 1/3, 0)$, $(1/3,2/3,1/3)$, and $(0,0,2/3)$ for the other (N or B, respectively). The rhombohedral structure has only two atoms per cell, located at $(0,0,0)$ and $(1/3,1/3,1/3)$. This point is crucial, since only six phonon modes are obtained, which is in contrast to the 18 modes of the ABC hexagonal structure. 

\subsection{Brillouin zones of the hexagonal and rhombohedral cells and band folding}

\begin{figure}[h]
\includegraphics[width=\columnwidth]{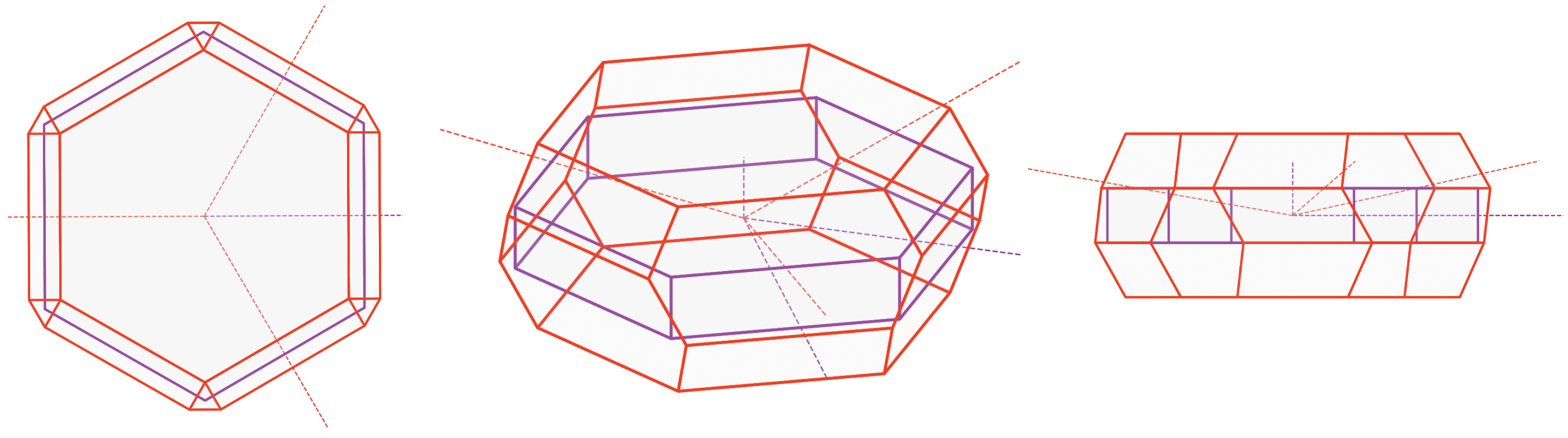}
\caption{\label{fig4SM} Brillouin zones of the hexagonal (purple lines) and rhombohedral (red lines) cells of the ABC stacking : top view (left), 3D view (center) and side view (right). Dotted lines represent the reciprocal vectors.}
\end{figure}

The Brillouin zones (BZ) of the rhombohedral and hexagonal lattices described previously are sketched in Fig.~\ref{fig4SM}. The BZ of the hexagonal cell is a prism with a hexagonal base and vertical facets, giving it a rotational symmetry of order $6$ in the $xy$-plane. The rhombohedral BZ is also based on a hexagon, but its facets are not vertical. They are composed of inclined rectangles and irregular hexagons. The hexagonal BZ is inscribed within the rhombohedral BZ, with a height equal to one third of the latter, as can be seen from the side view. 

\begin{figure}[h]
\includegraphics[width=0.45\columnwidth]{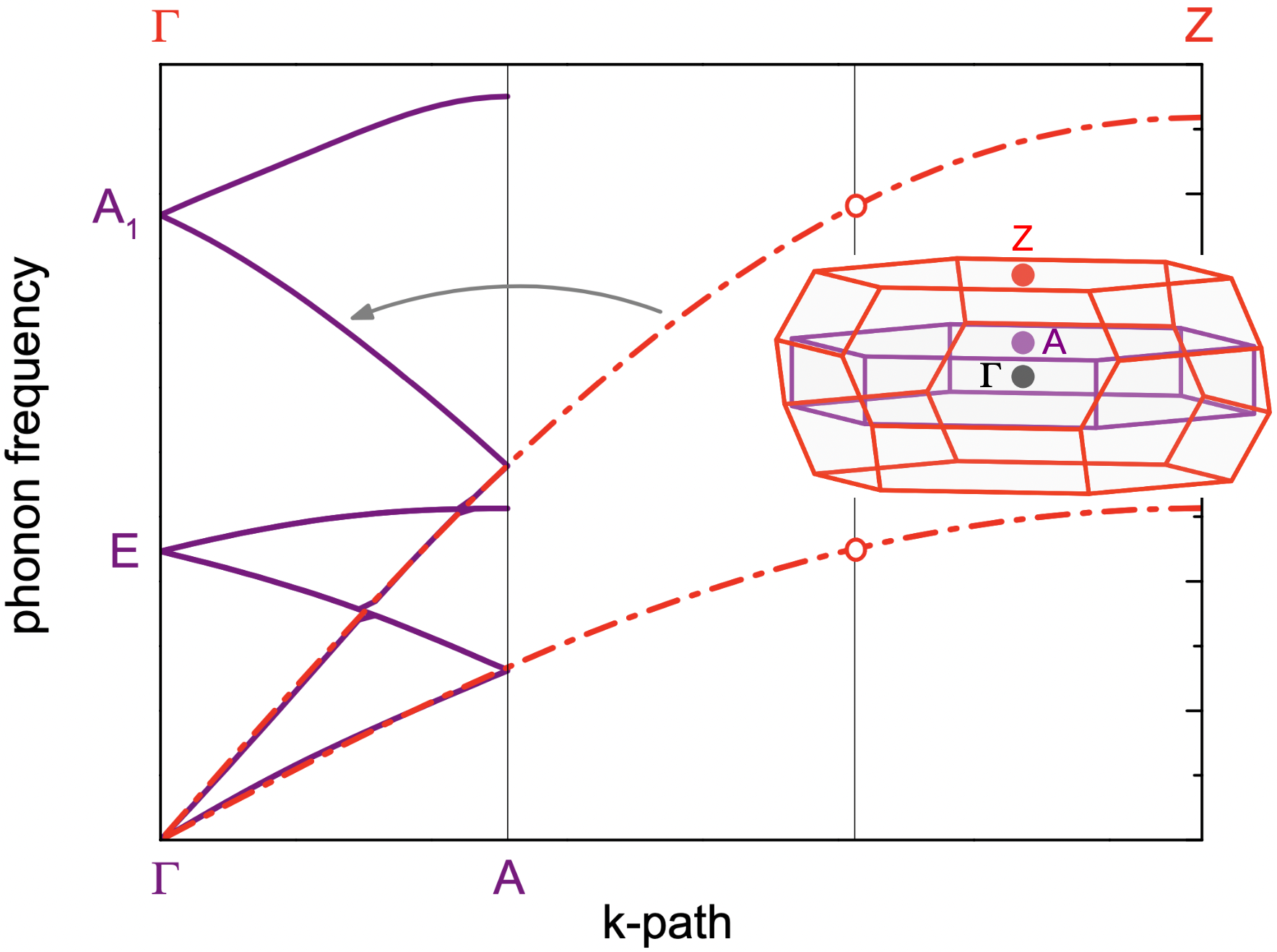}
\caption{\label{fig5SM} Acoustic phonon modes plotted along the $\Gamma A$ path (purple lines, bottom axis) for the hexagonal Brillouin zone of the ABC stacking and along the $\Gamma Z$ path (red lines, top axis) for the rhombohedral cell. The curved arrow represents the effect of band folding. The Brillouin zones and high symmetry points, $\Gamma$, $A$ and $Z$ are drawn in inset.}
\end{figure}

\begin{figure}[h]
\includegraphics[width=0.5\columnwidth]{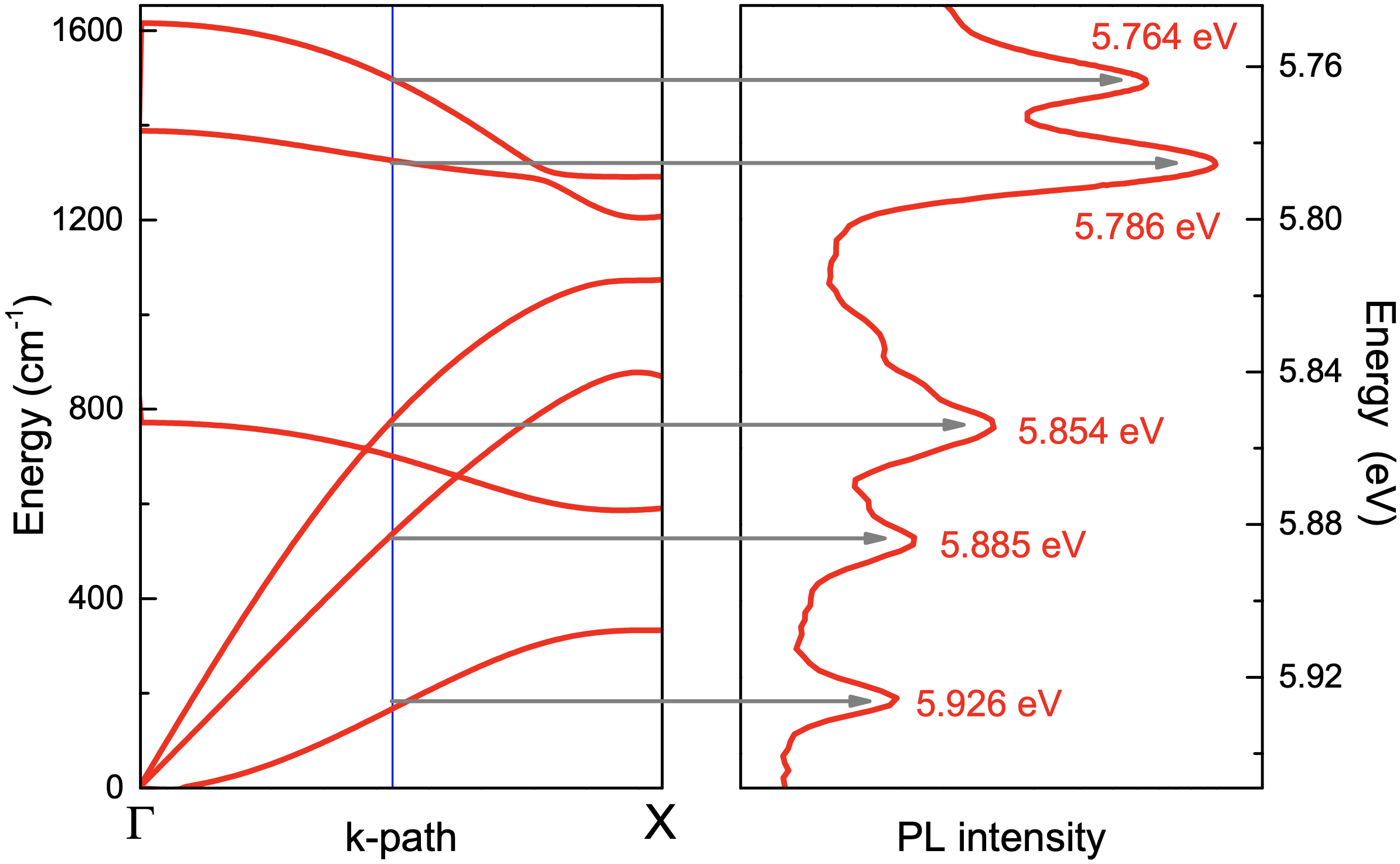}
\caption{\label{fig7SM} Phonon dispersion of rBN along the $\Gamma X$ path (left graph) and low temperature photoluminescence intensity plotted as a function of the energy on the vertical axis (right graph). The horizontal arrows highlight the correspondence of the phonon energies at $\frac{\Gamma X}{2}$ with the main PL peaks. The PL energy scale is offset by $5.949$~eV with respect to the phonon dispersion energy scale.}
\end{figure}

\begin{figure}[h]
\includegraphics[width=0.45\columnwidth]{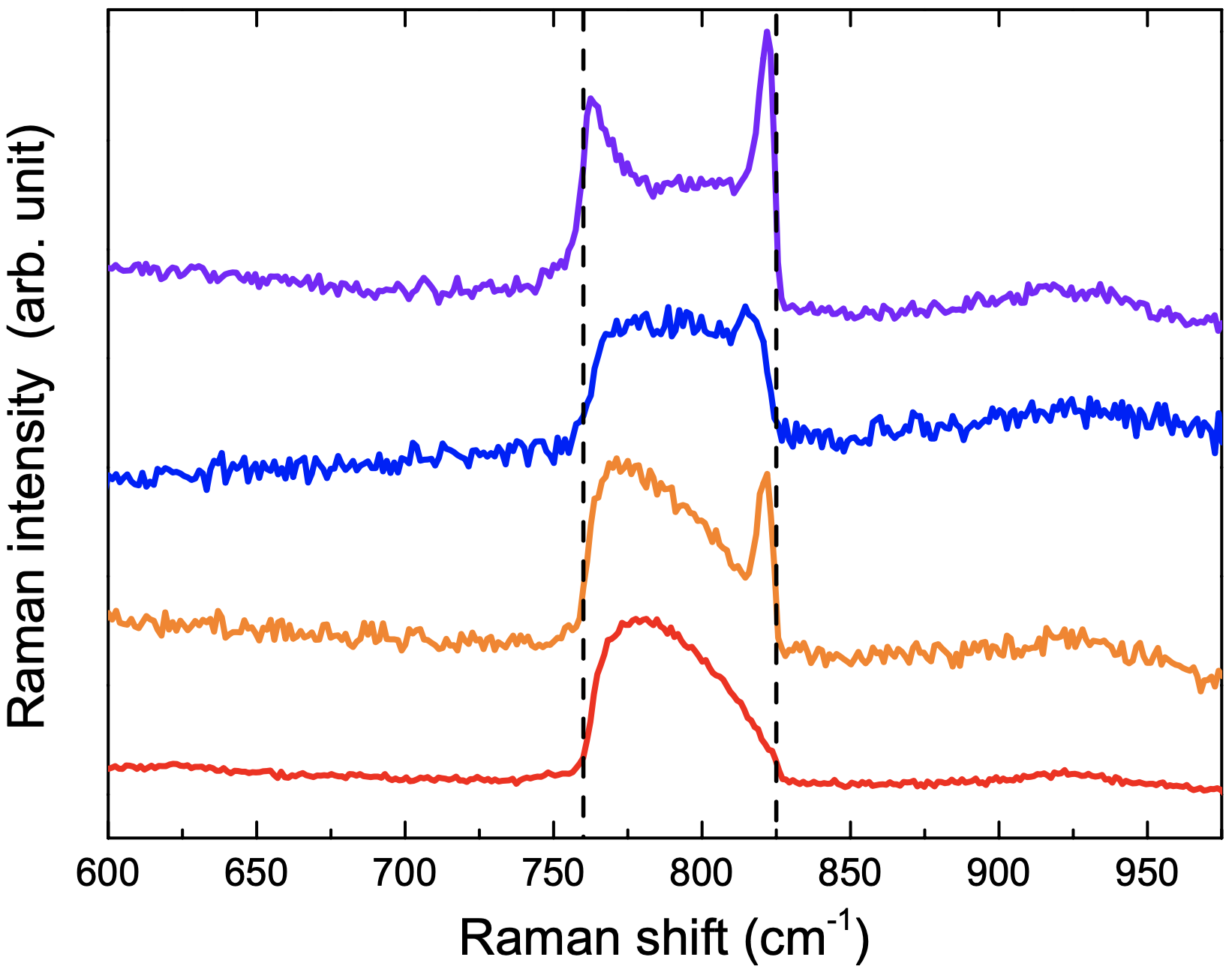}
\caption{\label{fig8SM} Different profiles of the extended band detected around $800$~cm$^{-1}$ on several rBN crystallites. The dashed vertical lines delimit the $760-825$~cm$^{-1}$ frequency band.}
\end{figure}

The location and labelling of the high-symmetry points differ for each BZ, meaning that the energy and phonon dispersions vs $k$-paths are not the same. To illustrate this, Fig.~\ref{fig5SM} plots the low-energy phonon modes along the $k_z$ axis, for $k_x=0$ and $k_y=0$. In the rhombohedral BZ (red lines, $\Gamma-Z$ path on the top axis), the double degenerate $E$ and $A_1$ modes appear as classical acoustic branches dispersing from $E=0$. In the hexagonal BZ representation (purple lines, $\Gamma-A$ path on the bottom axis), the previous modes fold twice, as highlighted by the gray arrow. This is because the $\Gamma A$ distance is one third that of $\Gamma Z$. This results in apparent non-zero phonon energies in $\Gamma$, i.e. $k_z=0$, which correspond to the energies of the point located at $k_z=2/3\;\Gamma Z$  in the rhombohedral representation (open circles in Fig.~\ref{fig5SM}). The absence of a low-energy mode at $\sim50$~cm$^{-1}$ in rBN can only be explained by considering the rhombohedral cell rather than the hexagonal cell of the ABC stacking.  
  
\subsection{Raman and photoluminescence supplementary data}

\end{document}